\documentclass[%
 reprint,
superscriptaddress,
 amsmath,amssymb,
]{revtex4-2}
\usepackage{siunitx}
\usepackage{graphicx}
\usepackage{dcolumn}
\usepackage{bm}

\usepackage{ulem}

\begin{document}

\preprint{APS/123-QED}

\title{Investigation of the SiO$_\mathrm{2}$-SiC interface using low energy muon spin rotation spectroscopy}

 \author{Piyush Kumar}
  \email[]{kumar@aps.ee.ethz.ch}
  \affiliation{Advanced Power Semiconductor Laboratory, Eidgenössische Technische Hochschule Z\"urich, Physikstrasse 3, 8092 Zurich, Switzerland}
 
 \author{Maria In{\^e}s Mendes Martins}%
\affiliation{Advanced Power Semiconductor Laboratory, Eidgenössische Technische Hochschule Z\"urich, Physikstrasse 3, 8092 Zurich, Switzerland}
\affiliation{Laboratory for Muon Spin Spectroscopy, Paul Scherrer Institute, Forschungsstrasse 111, 5232 Villigen
PSI, Switzerland}


 \author{Marianne Etzelmüller Bathen}
  \affiliation{Advanced Power Semiconductor Laboratory, Eidgenössische Technische Hochschule Z\"urich, Physikstrasse 3, 8092 Zurich, Switzerland}
 \author{Judith Woerle}
  \affiliation{Advanced Power Semiconductor Laboratory, Eidgenössische Technische Hochschule Z\"urich, Physikstrasse 3, 8092 Zurich, Switzerland}
\author{Thomas Prokscha}
\affiliation{Laboratory for Muon Spin Spectroscopy, Paul Scherrer Institute, Forschungsstrasse 111, 5232 Villigen
PSI, Switzerland}
 \author{Ulrike Grossner}
  \affiliation{Advanced Power Semiconductor Laboratory, Eidgenössische Technische Hochschule Z\"urich, Physikstrasse 3, 8092 Zurich, Switzerland}


\date{\today}

\begin{abstract}

Using positive muons as local probes implanted at low energy enables gathering information about the material of interest with nanometer depth resolution (low energy muon spin rotation spectroscopy~(LE-$\mu$SR)). In this work, we leverage the capabilities of LE-$\mu$SR to perform a detailed investigation of the widely studied yet poorly understood SiO$_2$-SiC interface. Thermally oxidized samples are investigated before and after annealing in nitric oxide~(NO) and argon~(Ar) environment. Thermal oxidation is found to result in structural changes both in the SiC crystal close to the interface and at the interface itself, which severely degrade the transport properties of carriers. Annealing in NO environment is known to passivate the defects leading to a reduction of the density of interface  traps~(D$_\mathrm{it}$); LE-$\mu$SR further reveals that the NO annealing results in a thin layer of high carrier concentration in SiC, extending to more than 50~nm depending on the annealing conditions. 
From our measurements, we see indications of Si vacancy (V$_\mathrm{Si}$) formation in SiC after thermal oxidation. Following NO annealing, nitrogen occupies the V$_\mathrm{Si}$ sites, leading to the reduction in D$_\mathrm{it}$ and, at the same time, creating a charge-carrier-rich region near the interface.
Thus, the LE-$\mu$SR technique sheds light on the near interface region in the SiO$_2$-SiC system which is challenging to access using other techniques. By comparing the LE-$\mu$SR data from a sample with known doping density, we perform a high-resolution quantification of the free carrier concentration near the interface after NO annealing and discuss the origin of the observed near-surface variations. 
Finally, the depletion of carriers in a MOS capacitor in the region exactly below the interface is demonstrated using LE-$\mu$SR. The NO annealed sample shows the narrowest depletion region, likely due to the reduced density of interface traps and charge-carrier-rich region near the interface. Our findings demonstrate the many benefits of utilizing LE-$\mu$SR to study critical regions of semiconductor devices that have been inaccessible with other techniques while retaining nanoscale depth resolution and a non-destructive approach.   
\end{abstract}

\maketitle


\section{\label{sec:level1}Introduction}
%
%
Muon spin rotation~($\mu$SR) spectroscopy is a powerful technique capable of studying intricate properties of materials from various disciplines such as 
superconductivity, magnetism, semiconductor defect characterization,
physical chemistry and dynamics of soft matter \cite{MuonSpectroscopyStephen,hillierMuonSpinSpectroscopy2022,pattersonMuoniumStatesSemiconductors1988,cox_muonium_2009,mckenziePositiveMuonMSR2013}.
%
%
Positive muons~($\mu^+$) act as microscopic probes and gather information about the region where they are implanted in the material. For magnetic materials, the $\mu$SR method has been successful in extracting the magnetic order parameter and magnetic volume fractions of the system and studying the dynamics of the magnetic moments in a sample \cite{hiraishi_bipartite_2014,frandsen_volume-wise_2016,prattMuonSpinRelaxation2009}, whereas for various unconventional superconductors, it has been used to understand the relation and coexistence between magnetism and superconductivity \cite{drewCoexistenceStaticMagnetism2009,luetkens_electronic_2009}.
$\mu$SR also finds special applications in the field of semiconductors, where muons can replicate the behaviour of hydrogen related defects which otherwise are challenging to probe \cite{hillierMuonSpinSpectroscopy2022,pattersonMuoniumStatesSemiconductors1988,cox_muonium_2009}. \\
A recent study with low energy~(LE)-$\mu$SR on germanium showed that charge carrier profiles can be determined in the accumulation and depletion regions near the surface \cite{prokschaDirectObservationHole2020a}, whereas in silicon carbide~(SiC), the  sensitivity of the $\mu$SR signal to defects created by proton irradiation has been demonstrated \cite{woerle2019interaction,woerle2020muon}. The sensitivity of $\mu$SR to important semiconductor parameters combined with the nano-scale resolution of the technique mark its low energy mode as an intriguing choice for investigating device related interface and surface regions. \\ 
4H-SiC is among the most prominent and well known semiconducting materials in the world of power semiconductor devices. Having a wide bandgap, high thermal conductivity and high saturation velocity, SiC holds the key to creating efficient high power devices \cite{choykePhysicalPropertiesSiC1997}. With the research moving swiftly from silicon to other competing materials, SiC offers the advantage that it is the only compound semiconductor with SiO$_\mathrm{2}$ being its native oxide \cite{Kimoto_2014}. In addition to being thermally stable and possessing high permittivity, SiO$_\mathrm{2}$ is especially useful as it would allow SiC devices to be fabricated with minor modifications to the already functional silicon~(Si) fabrication facilities.
However, the concentration of thermal oxidation induced defects are at least one order of magnitude higher for SiC than its Si counterpart. These defects are present at and near the interface and can have a  debilitating impact on the carrier transport, leading to a high channel resistance. For a medium power device such as a 1.2~kV SiC power MOSFET, the channel resistance can be as high as \SI{60}{\percent} of the total on-state resistance \cite{muting2018simulation}. Although annealing in a NO environment is known to reduce the density of interface traps~(D$_\mathrm{it}$) \cite{dhar2005interface}, neither the defects formed by the initial oxidation and their distribution, the role played by nitrogen during annealing, nor the consequent changes at the interface are fully understood. 

\begin{table*}[t]
\caption{\label{tab:table1} Description of sample preparation and processing steps. Deposition was performed using PE-CVD for SiO$_\mathrm{2}$ and PVD for Al. }
\begin{ruledtabular}
\begin{tabular}{ccccc}
\textbf{Sample name} & \textbf{Oxidation parameters} &\textbf{ Oxide thickness} & \textbf{POA parameters} & \textbf{Deposition}\\
\hline
1300x & 32 min at 1300~$^\circ$C & 48~nm & - & SiO$_\mathrm{2}$, 52~nm \\
1300NO1150 & 30 min at 1300~$^\circ$C &  48~nm & 70 min at 1150~$^\circ$C in NO  & SiO$_\mathrm{2}$, 52~nm\\
1300NO1300  & 28 min at 1300~$^\circ$C&  56~nm & 70 min at 1300~$^\circ$C in NO  & SiO$_\mathrm{2}$, 44~nm\\
1300Ar1300  & 32 min at 1300~$^\circ$C& 48~nm & 70 min at 1300~$^\circ$C in Ar  & SiO$_\mathrm{2}$, 52~nm\\
\hline
1300x-Al & 32 min at 1300~$^\circ$C& 48~nm & - & Al, 52~nm\\
1300NO1150-Al & 30 min at 1300~$^\circ$C&  48~nm & 70 min at 1150~$^\circ$C in NO  & Al, 52~nm\\
1300NO1300-Al  & 28 min at 1300~$^\circ$C&  56~nm & 70 min at 1300~$^\circ$C in NO  & Al, 44~nm\\
1300Ar1300-Al  & 32 min at 1300~$^\circ$C& 48~nm & 70 min at 1300~$^\circ$C in Ar  & Al, 52~nm\\

\end{tabular}
\end{ruledtabular}
\end{table*}

\noindent Herein, we exploit the strength of LE-$\mu$SR to study the impact of thermal oxidation and post oxidation annealing~(POA) in a NO environment on the SiO$_\mathrm{2}$-SiC interface in a depth-resolved manner. A detailed quantification of the carrier concentration near the surface of the NO annealed samples is performed, revealing an unintentional doping during the annealing and the variation of carrier concentration from the interface towards the SiC bulk. Further, a thin layer of metal is deposited on the oxide to fabricate a metal-oxide-semiconductor~(MOS) capacitor, and the resulting changes in carrier concentration due to the potential variation at the surface of the semiconductor are studied using LE-$\mu$SR. 

\section{Methodology}
For the experiments, n-type, nitrogen doped 4H-SiC samples with \SI{30}{\micro\meter} thick epitaxial layers were employed. Each sample in the set underwent thermal oxidation and two of the samples were exposed to a POA treatment in NO ambient, and one sample in Ar ambient. The doping density of each sample is $\sim$\SI{2.6e15}{\per\centi\meter\cubed}. A thin layer of SiO$_\mathrm{2}$ was deposited (using plasma enhanced chemical vapour deposition~(PE-CVD)) on top of the thermal oxide to reach a total thickness of 100~nm, such that the SiC-SiO$_\mathrm{2}$ interface lies at the same distance from the surface for each sample. In a second set, a thin layer of aluminum~(Al) was deposited using physical vapour deposition~(PVD) on top of the thermal oxide instead of the PE-CVD oxide to reach a total thickness of 100~nm. An overview of the sample processing parameters is presented in Table~\ref{tab:table1}. \\
The samples have been characterized with low-energy muon spin rotation spectroscopy~(LE-$\mu$SR), capacitance-voltage~(CV) method and deep level transient spectroscopy~(DLTS). 
A theoretical study was performed using density functional theory (DFT) calculations to investigate the preferred position of nitrogen in the 4H-SiC lattice and its proclivity for motion under the relevant annealing conditions. 

\subsection{LE-$\mu$SR}

\begin{figure*}
	\centering
	\includegraphics{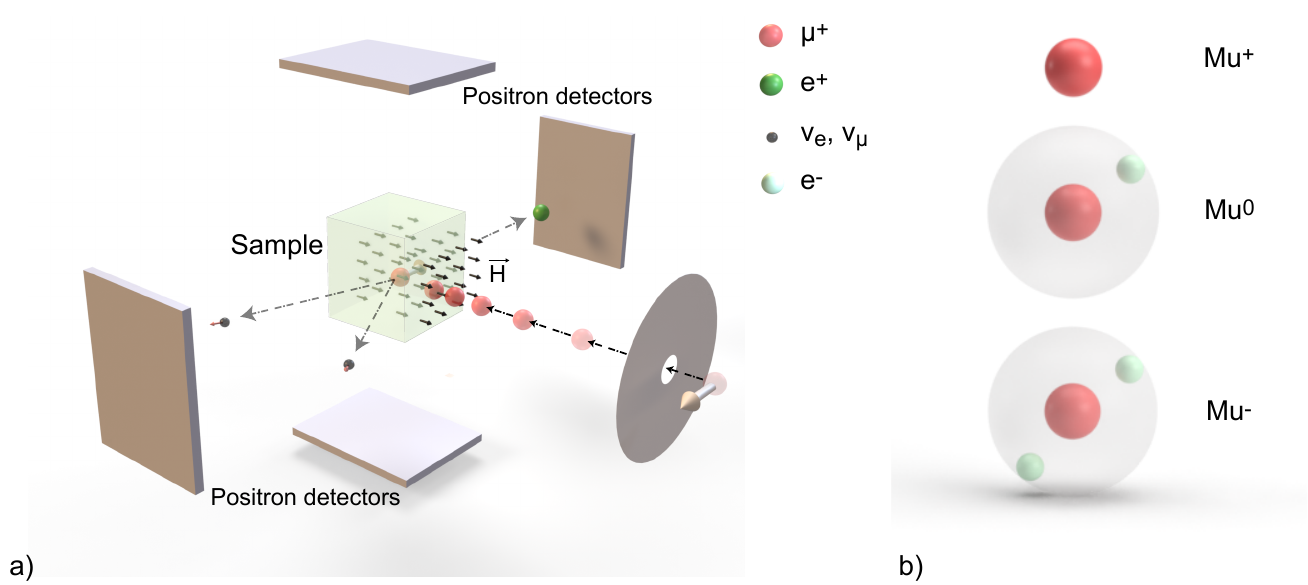}
	\caption{a) Illustration of the $\mu$SR experimental setup. Spin polarised muons with their spins pointing perpendicular to the direction of motion are implanted in the sample. Under the influence of the local and the applied external magnetic fields, the muon precesses and decays with a mean lifetime of \SI{2.2}{\micro\second} into a positron and two neutrinos. The decay positrons are detected using the positron detectors. b) The three possible states of muonium inside the semiconductor.   }
	\label{fig:uSR setup}
\end{figure*}

\begin{figure}
	\centering
	\includegraphics{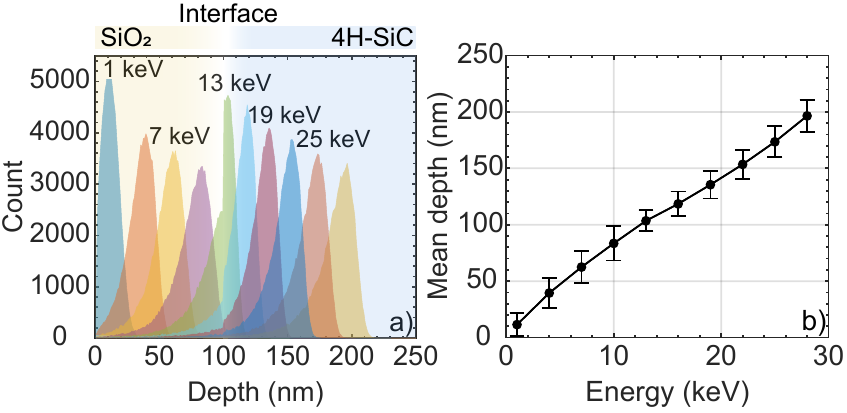}
	\caption{a) Muon implantation profile for energies between 1 and 28~keV, simulated with TRIMSP \cite{Eckstein_1991,Morenzoni_2002}. b) The calculated mean implantation depth for each implantation energy. }
	\label{fig:muonImplantation}
\end{figure}

Positively charged muons~($\mu^+$) used in a $\mu$SR experiment are almost 100~\% spin~($S = 1/2$) polarised. They decay into a positron and two neutrinos with a mean lifetime of \SI{2.2}{\micro\second}.
The positron is emitted preferentially in the direction of the muon spin at the time of decay and carries the information about the time-evolution of the muon spin polarisation under the influence of the local and external magnetic fields. The LE-$\mu$SR experimental setup is illustrated in Fig.~\ref{fig:uSR setup}~a). \\
In a conventional $\mu$SR experiment, the $\mu^+$ is implanted at a high energy of around 4~MeV which leads to a penetration depth of hundreds of \SI{}{\micro\meter} to a few \SI{}{\milli\meter} into the material. In order to study the SiO$_\mathrm{2}$-SiC interface, the $\mu^+$ are slowed down using a moderator \cite{prokschaModeratorGratingsGeneration2001} and then re-accelerated to energies from 1-28~keV. This energy range translates to tunable mean implantation depths between 10~and~150~nm in SiC. Using low energy muons a detailed depth profile can be obtained for the interface and the near interface regions of the  oxide-semiconductor structure. Muon implantation profiles in the SiO$_\mathrm{2}$-SiC system are shown in Fig.~\ref{fig:muonImplantation}. The mean implantation depth
%
%
is extracted from these profiles and is used 
throughout the rest of this manuscript. \\

\noindent When the $\mu^+$ is implanted in the material it undergoes inelastic scattering processes where it is decelerated and creates a track of ionization electrons. 
As the energy of the $\mu^+$ is reduced, fewer electrons are ionized and below a few keV, no further ionization track electrons are generated. The $\mu^+$ then loses its energy by charge-exchange cycles. 
The final state of the muon can be the neutral muonium state  Mu$^0$, a hydrogen-like bound state of a $\mu^+$ with an e$^-$.
Muonium  behaves like a light isotope of hydrogen inside the semiconductor and
can exist in three differently charged configurations: the paramagnetic neutral Mu$^{0}$, and the charged diamagnetic negative (Mu$^{-}$) or positive (Mu$^{+}$) muonium configurations.  
\\ 
In a hole rich environment, the Mu$^0$ can capture a hole to form Mu$^+$, whereas in an electron rich environment, Mu$^0$ can capture an electron to form Mu$^-$. The three charge states of the muonium atom are illustrated in Fig.~\ref{fig:uSR setup}~b). \\
Due to the hyperfine interaction between the bound $\mathrm{e}^-$ and the $\mu^+$, the precession frequency of the muon spin in the paramagnetic Mu$^0$ in a transverse magnetic field $B$ is considerably different to the Larmor precession frequency $\omega_D = \gamma_D\cdot B$ of the muon spin in its diamagnetic states~($\gamma_D$ is the gyromagnetic ratio of the muon). In low transverse magnetic fields~($B \leq$~1~mT) the transition frequencies between the triplet states in Mu$^0$ with isotropic hyperfine coupling are degenerate. Here, the triplet Mu$^0$ precession
can be observed at a frequency  $\omega_\mathrm{Mu} = \gamma_\mathrm{Mu}\cdot B$, where the gyromagnetic ratio $\gamma_\mathrm{Mu}$ of the triplet state is about 103 times larger than the gyromagnetic ratio $\gamma_{D}$ of the muon.
%
%
At higher fields, the transition frequencies between the triplet states split \cite{MuonSpectroscopyStephen}. 
The precession frequencies of Mu$^0$ at 
$B > 4$~mT cannot be resolved by the spectrometer at the LE-$\mu$SR setup and appear as a loss of signal. Therefore, all measurements in this work were performed at a low field of 0.5~mT.
\\ 
The large difference between the precession frequencies of Mu$^0$ and Mu$^+$/Mu$^-$ allows to readily resolve the paramagnetic and the diamagnetic components. In low transverse magnetic fields with degenerate triplet Mu$^0$ precession frequencies, the 
%
%
muon decay asymmetry $A(t)$ is given by:
%
\begin{equation}
\begin{split}
         A(t)=A_\text{D}  \exp(-\lambda_\text{D}  t)  \cos(\omega_\text{D} t + \phi_\text{D})+\\
     A_\text{Mu}  \exp(-\lambda_\text{Mu} t) \cos(\omega_\text{Mu} t + \phi_\text{Mu}),
\end{split}
\label{eqn: AssymetryEquation}
\end{equation}
%
where A$_\mathrm{D}$ and A$_\mathrm{Mu}$ are the decay asymmetries of the diamagnetic and the paramagnetic states with precession frequencies $\omega_\mathrm{D}$ and $\omega_\mathrm{Mu}$, respectively, and $\lambda_\mathrm{D}$ and $\lambda_\mathrm{Mu}$ are the corresponding depolarisation rates. 
The initial phases $\phi_\text{D}$ and $\phi_\text{Mu}$ of the diamagnetic and paramagnetic $\mu$SR signals are determined by the position of a specific positron detector with respect to the initial muon spin direction. If the diamagnetic state forms from a neutral
precursor state -- where the muon spin moves in the opposite direction to the Larmor precession -- a negative phase shift of $\phi_\text{D}$ with respect to the muon Larmor precession can be observed for the diamagnetic
signal, if the transition rate to the diamagnetic state is on the order of hundreds of MHz  \cite{pattersonMuoniumStatesSemiconductors1988,prokschaDirectObservationHole2020a}.
%
The longer the muon stays in the precursor Mu$^0$ state, the more negative the phase becomes. Within a semiconductor sample, narrow regions with different carrier concentrations 
strongly impact the measured asymmetry and phase. 
In n-type regions of a semiconductor, the Mu$^0$ state can capture a free electron to form the diamagnetic Mu$^-$ state. The higher the density of free electrons, the faster is the capture process, resulting in an increase of the measured $A_\mathrm{D}$ and a corresponding change of the negative phase shift as a function of free carrier density.
%
Therefore, the diamagnetic asymmetry and phase of the $\mu$SR signal strongly depend on the carrier concentration. 
%
%
%
\subsection{Electrical characterization}
Thermal oxidation of SiC is  known to create electrically active defects at and near the interface. These defects are commonly studied with C-V analysis to gauge their concentration and energy levels in the band gap  \cite{yoshiokaAccurateEvaluationInterface2012a}. Circular 200~nm thick Aluminium~(Al) contacts of diameter varying between \SI{200}{\micro\meter} to \SI{600}{\micro\meter} were deposited on top of the thermal oxide using an e-beam evaporator to create a metal oxide semiconductor~(MOS) structure. A 100~nm thick layer of Ni was deposited on the C-face to get a good ohmic contact. Quasi static and high frequency~(2~MHz) C-V measurements were  performed on the MOS capacitors and used to calculate the density of interface traps~(D$_\mathrm{it}$). 
\\
The samples were also characterised using DLTS using a reverse bias of 0~V and a pulse voltage of -5~V. A pulse width of 100~ms and period width of 500~ms are used for each measurement. The DLTS signal refers to the coefficient of the sine term~(b1) in the Fourier series of the deep level transient Fourier spectroscopy~(DLTFS)~\cite{weissDeepLevelTransient1988}. 
In the experiment, we have focused on the N-related defects in the SiC and have therefore chosen a temperature range from \SIrange{25}{60}{\kelvin}.

\subsection{Density functional theory calculations}
Density functional theory (DFT) calculations as implemented in the Vienna ab initio simulation package or {\scriptsize VASP} code  \cite{Kresse1996,Kresse1996a} were performed to investigate defect energetics and kinetics for nitrogen incorporation into the 4H-SiC lattice. The lattice parameters were obtained by relaxing the 4H-SiC unit cell using the hybrid Heyd-Scuseria-Ernzerhof HSE06 functional \cite{Heyd2003} until the forces were below \SI{0.005}{\electronvolt\per\angstrom}, yielding a band gap of $3.17$~eV close to the experimental value. 400-atom and 576-atom 4H-SiC supercells were constructed from the relaxed unit cell and defects created by adding and removing atoms. 
Defect relaxations were performed using 576-atom supercell, the GGA-level Perdew-Burke-Ernzerhof (PBE) functional \cite{Perdew1996} and a $2\times2\times2$ \textbf{k}-mesh. 
The convergence criterion for the electronic self-consistent loop was set to \SI{1e-6}{\electronvolt} for a geometric relaxation criterion of \SI{0.01}{\electronvolt\per\angstrom} maximum forces and a \SI{450}{\electronvolt} energy cut-off. 
Defect energetics were obtained self-consistently by single-shot calculations for the same supercell and parameters using the HSE06 functional and $\Gamma$-only \textbf{k}-point sampling.  

Formation energy diagrams for point defects were constructed according to Ref.~\cite{Freysoldt_2014}: 
\begin{equation}
E^\mathrm{f}(q) = E^\mathrm{total}_\mathrm{defect}(q) - E^\mathrm{total}_\mathrm{supercell}-\Sigma_i n_i \mu_i +q(E_\mathrm{V}+E_\mathrm{F})+E^\mathrm{FNV},    
\end{equation}
where $E^\mathrm{total}_\mathrm{defect/supercell}$ refers to the total energy of the defective and pristine supercells, respectively, $n_i$ is the number of atoms added ($n_i>0$) or removed ($n_i<0$) from the supercell to create the defect, and $\mu_i$ is the chemical potential for each species (Si, C, N). 
Formation energy diagrams were computed in the Si poor/C rich limit. Chemical potentials for Si and C were determined from the total energy per atom of the unit cell of Si and diamond, respectively, while $\mu_\mathrm{N}$ was obtained from Si$_\mathrm{3}$N$_\mathrm{4}$.  
The unit cells to obtain $\mu_i$ for Si, C and N were relaxed using HSE06, $700$~eV energy cutoff, a \textbf{k}-mesh density of 4, and for a geometric relaxation criterion of \SI{0.005}{\electronvolt\per\angstrom} maximum forces. 
Furthermore, $E_\mathrm{V}$ refers to the valence band maximum and $E_\mathrm{F}$ is the Fermi level position relative to $E_\mathrm{V}$. 
The total energies for charged defects were corrected according to the Freysoldt-Neugebauer-van de Walle (FNV) correction scheme \cite{Freysoldt_2009} where $E^\mathrm{FNV}$ is a
correction term to account for the use of charged and finite-sized supercells. 

Defect migration was studied using the nudged elastic band (NEB) method \cite{JONSSON1998}. NEB calculations take as input two fixed structures: the initial and finite state of the migration. A chain of intermediate structures, or 'images', is then formed as a first guess along the migration path. The images are connected via spring forces to keep them sufficiently distinct from one another and optimized collectively. The NEB calculations herein were performed for interstitial nitrogen entering the Si vacancy in neutral and negative charge state using 5 intermediate images, a $400$~eV energy cutoff, $\Gamma$-only \textbf{k}-point sampling and the PBE functional. The forces were relaxed until the maximum force was below \SI{0.05}{\electronvolt\per\angstrom} for the migration paths.  

\section{\label{sec:RnD}Results and discussion}

The two sets of samples described in Table~\ref{tab:table1} were studied with LE-$\mu$SR. The experiments have been performed at a low magnetic field of 0.5~mT and at 10~K and 260~K. By combining low and high temperature measurements the impact of charge carriers on the $\mu$SR signal can be clearly distinguished. 

\subsection{LE-$\mu$SR study of oxide-semiconductor samples}

\begin{figure}
	\centering
	\includegraphics{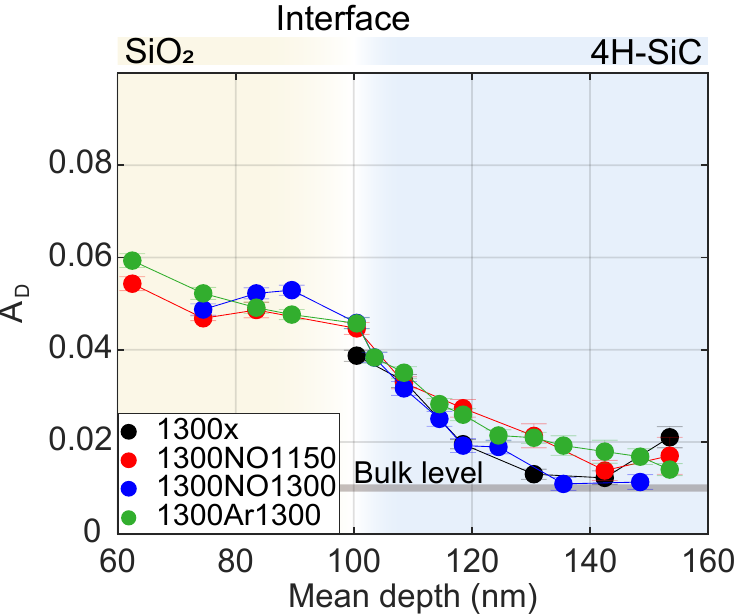}
	\caption{Diamagnetic asymmetry $A_\text{D}$ as a function of mean depth 
	measured at 10~K and 0.5~mT. 
	}
	\label{fig:10K_5G}
\end{figure}

Fig.~\ref{fig:10K_5G} compares the LE-$\mu$SR data at 10~K for thermally oxidised samples that experienced different POA conditions. The light yellow region represents the thermally grown SiO$_\mathrm{2}$  whereas the blue region is the 4H-SiC epi-layer. The shaded region in-between represents the oxide-semiconductor interface.
Typically, in thermally grown SiO$_\mathrm{2}$, Mu$^0$ formation is supported due to high structural order and therefore a small diamagnetic asymmetry (A$_\mathrm{D}$) signal is observed. 
However, close to the interface a high concentration of defects or structural disorder is expected which leads to an increase in A$_\mathrm{D}$ at the expense of the paramagnetic Mu$^0$~\cite{martins_defect_2022}. In 4H-SiC, at 10~K, a freeze out of carriers is expected. 
Due to the lack of carriers the implanted muon has a low probability to form the diamagnetic states Mu$^+$ or Mu$^-$. The diamagnetic asymmetry therefore drops to even smaller values after the interface and only slight differences in the muon behavior can be observed across the four samples.

\begin{figure}
	\centering
	\includegraphics{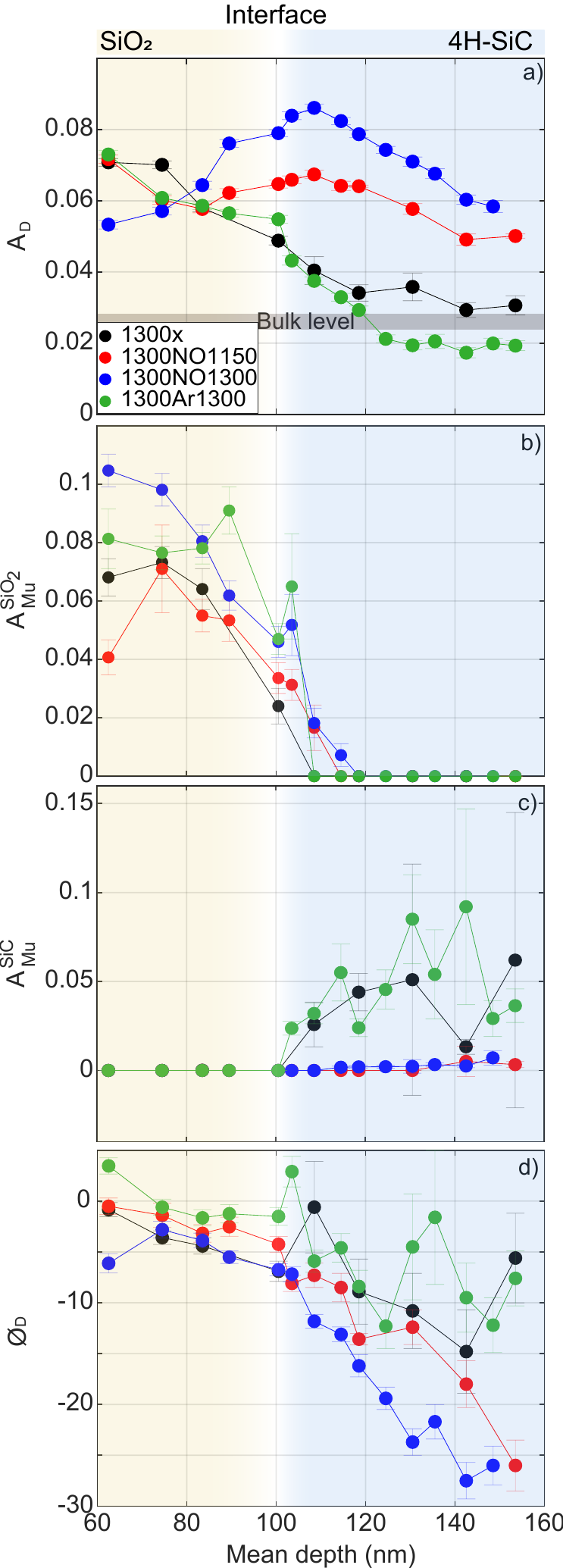}
	\caption{LE-$\mu$SR signals recorded at 260~K and 0.5~mT. a) Diamagnetic asymmetry~(A$_\text{D}$) as a function of mean depth. 
	Paramagnetic asymmetry for the Mu$^0$ precessing at b) $\omega_\mathrm{Mu}^{\mathrm{SiO_2}}$ and  c) $\omega_\mathrm{Mu}^{\mathrm{SiC}_a}$, $\omega_\mathrm{Mu}^{\mathrm{SiC}_b}$. 
	The sharp drop of the A$_\mathrm{Mu}^{\mathrm{SiO_2}}$ and a rise of A$_\mathrm{Mu}^\mathrm{{SiC}}$~(c) at the SiO$_2$-SiC interface shows that the oxide semiconductor interface can be identified. 
	Almost no paramagnetic signal is observed for the NO annealed samples, indicating a Mu$^0$ to Mu$^-$ transformation.  d) Phase $\phi_\text{D}$.
	} 
	\label{fig:260K_5G}
\end{figure}

\noindent The $\mu$SR data recorded at 260~K is shown in Fig.~\ref{fig:260K_5G}. The  A$_\mathrm D$ is no longer similar for the four samples that underwent different POA procedures. Instead, substantial differences across the set are observed, in addition to remarkable changes compared to the 10~K case. The slight increase in diamagnetic asymmetry for the non-annealed sample (1300x) compared to the 10~K data can be attributed to the ionization of the nitrogen dopants at 260~K. 
Moreover, a remarkable increase can be observed for the NO annealed samples compared to the data recorded at 10~K and at 260~K for the samples that did not receive a POA in NO (1300x, 1300Ar1300). Since all the samples have very similar initial doping densities the semiconductor doping itself cannot explain the large difference observed in the diamagnetic signals in Fig.~\ref{fig:260K_5G}~a). 
\\
The bulk level shown in Fig.~\ref{fig:10K_5G} and \ref{fig:260K_5G}~a) as reference is taken from the $\mu$SR data of an unprocessed, low doped n-type 4H-SiC sample \cite{martins_defect_2022}. The POA in NO environment has a substantial impact on the diamagnetic signal. Looking closely at the 1300NO1300 sample, the diamagnetic value is high at the interface and then begins to drop, still staying higher than the non-annealed sample at the end of the probing range around 55~nm from the interface. In other words, the changes imposed on the SiC matrix by the NO annealing remain prominent far beyond the point where the $\mu$SR signal for the 1300x sample reaches the bulk level. On the contrary, annealing in an Ar environment results in a significant drop in the A$_\mathrm{D}$ signal to below the bulk level. The A$_\mathrm{D}$ value for the Ar annealed sample at 260~K is similar to that at 10~K, indicating similar carrier concentration in the first 60~nm of SiC in both cases. 
An important conclusion that can be drawn by looking at the A$_\mathrm{D}$ value for the NO and Ar annealed samples is that the increase in the A$_\mathrm{D}$ signal for NO annealed samples cannot be attributed to the temperature alone and NO has a pivotal role to play at the oxide-semiconductor interface.
\\
The paramagnetic Mu$^0$ has a single precession frequency in SiO$_\mathrm{2}$~($\omega_\mathrm{Mu}^{\mathrm{SiO}_\mathrm{2}}$) while it splits into two lines~($\omega_\mathrm{Mu}^{\mathrm{SiC}_a}$, $\omega_\mathrm{Mu}^{\mathrm{SiC}_b}$) due to an anisotropic hyperfine coupling in 4H-SiC \cite{martins_defect_2022}. Due to the difference in the precession frequencies, the 
muonium states in  SiO$_\mathrm{2}$ and SiC
can be distinguished, enabling identification of the interface location.  Fig.~\ref{fig:260K_5G}~b) shows the paramagnetic signal recorded in SiO$_2$~(A$_\mathrm{Mu}^{\mathrm{SiO}_\mathrm{2}}$) precessing at $\omega_\mathrm{Mu}^{\mathrm{SiO_2}}$. The signal reduces upon approaching the interface and drops to zero inside the semiconductor. Fig.~\ref{fig:260K_5G}~c) shows the paramagnetic asymmetry~(A$_\mathrm{Mu}^{\mathrm{SiC}}$) recorded for the Mu$^0$ inside the SiC. This signal is absent in the SiO$_\mathrm{2}$ and rises after the SiO$_\mathrm{2}$-SiC interface. Using the paramagnetic signals A$_\mathrm{Mu}^{\mathrm{SiC}}$ and A$_\mathrm{Mu}^{\mathrm{SiO}_\mathrm{2}}$, the two regions of the oxide-semiconductor sample can be clearly identified, the interface can be located and the thickness of the oxide can be estimated. In case of the non-annealed and Ar annealed samples, a sharp rise of the A$_\mathrm{Mu}^{\mathrm{SiC}}$ signal~(Fig.~\ref{fig:260K_5G}~c) at a mean depth of 100~nm confirms the position of the interface and consequently the thickness of the oxide.
\\
\\
The dependence of the paramagnetic asymmetry on the annealing conditions follows a different pattern in the SiC region as demonstrated by the  A$_\mathrm{Mu}^{\mathrm{SiC}}$ signal in Fig.~\ref{fig:260K_5G}~c). The 1300x and 1300Ar1300 samples exhibit similar behavior while the contrast is stark to the case after NO annealing. 
For the NO annealed samples almost no Mu$^0$ precession is observed, indicating that the fraction of the implanted muons forming Mu$^0$ quickly transforms to Mu$^-$ or quickly depolarizes due to the interaction with free charge carriers \cite{pattersonMuoniumStatesSemiconductors1988,martins_defect_2022}. A sharp reduction in phase was recorded, see Fig.~\ref{fig:260K_5G}~d), indicating that the muon transformed to the diamagnetic state from a neutral precursor state \cite{prokschaDirectObservationHole2020a}. \\
The changes in muon response as a function of annealing conditions have two potential causes: (i) differences in defect distribution before and after POA, and (ii) differences in carrier concentration.

In order to access the sensitivity of $\mu$SR to the defects in 4H-SiC, Woerle~${et~al.}$~\cite{woerle2019interaction,woerle2020muon} 
showed that the presence of carbon vacancies~(V$_\mathrm{C}$) in n-type 4H-SiC leads to an observable increase in the diamagnetic signal due to the formation of Mu$^-$ by electron capture of a neutral precursor state at a carbon vacancy, if the V$_\mathrm{C}$ concentrations $n_{Vc}> 10^{17}$~cm$^{-3}$, while the Mu$^0$ precession signal remains observable for $n_{Vc} \sim 10^{18}$~cm$^{-3}$.
%
However, thermal oxidation is known to reduce the V$_\mathrm{C}$ concentration in SiC and it was shown that the carbon interstitials generated during thermal oxidation can diffuse into the bulk and passivate this lifetime limiting defect \cite{hiyoshi2009reduction,kawahara2012analytical}. Therefore, the observed increase in the A$_\mathrm{D}$ signal after NO annealing cannot be attributed to the presence of V$_\mathrm{C}$. 
The presence of silicon vacancies~(V$_\mathrm{Si}$) in 4H-SiC has the opposite impact on the diamagnetic signal \cite{woerle2020muon}. Ab initio calculations revealed that the V$_\mathrm{Si}$-Mu$^0$ complex is energetically more favourable than isolated V$_\mathrm{Si}$ and leads to a reduction of A$_\mathrm{D}$. V$_\mathrm{Si}$ has been theorized to be present at and near the SiC-SiO$_\mathrm{2}$ interface after thermal oxidation. With spin dependent recombination~(SDR) studies, signatures were found suggesting that the V$_\mathrm{Si}$ is an important  contributor to the density of interface traps~(D$_\mathrm{it}$) \cite{cochraneIdentificationSiliconVacancy2012, cochrane2013effect}. During the thermal oxidation process, Si from the SiC matrix is expected to diffuse outwards and react with the incoming oxygen to form the oxide layer \cite{hijikataThermalOxidationMechanism2012,hijikataMacroscopicSimulationsSiC2019}, pointing towards the presence of V$_\mathrm{Si}$ near the interface. Considering Fig.~\ref{fig:260K_5G}~a), a drop in the A$_\mathrm{D}$ signal to the bulk level for 1300x and below the bulk level for 1300Ar1300 indicate a high concentration of V$_\mathrm{Si}$ near the interface for the non-annealed and Ar-annealed samples. 

\begin{figure}
	\centering
	\includegraphics{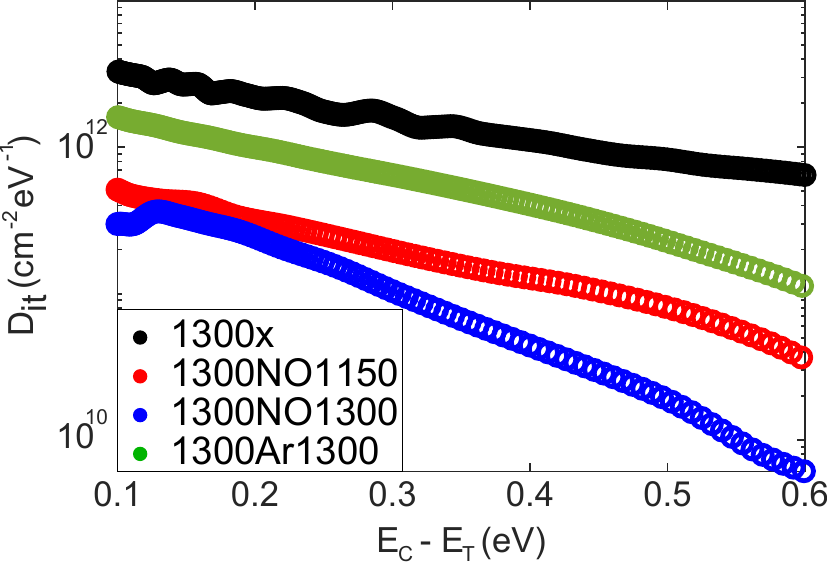}
	\caption{Density of interface traps $\text{D}_\text{it}$ extracted using the C-V method. E$_\text{C}$ is the conduction band edge and E$_\text{T}$ is the energetic position of the trap level in the band gap. The non-annealed sample (1300x) has a significantly higher D$_\mathrm{it}$ when compared to the NO annealed samples (1300NO1150 and 1300NO1300).}
	\label{fig:Dit}
\end{figure}

POA in NO environment is known to significantly reduce the D$_\mathrm{it}$ \cite{dhar2005interface} when compared to POA in an Ar environment. The D$_\mathrm{it}$ measured for the four samples are shown in Fig.~\ref{fig:Dit} and exhibit the expected behavior. Assuming that the V$_\mathrm{Si}$ created by thermal oxidation do indeed have a significant contribution to the D$_\mathrm{it}$, a reduction in the D$_\mathrm{it}$ after POA in the NO environment means that the nitrogen travels through the oxide and passivates the defects at and near the interface.
Additionally, nitrogen is likely incorporated into the SiC matrix resulting in a thin region of high carrier concentration \cite{liuEnhancedInversionMobility2013}. Scanning capacitance microscopy revealed doping levels of around $5\times10^{17}$~cm$^{-3}$ for samples annealed in N$_2$O \cite{fiorenzaSiO4HSiCInterface2013} in the region close to the interface. The bulk doping level for the SiC was $\sim$\SI{5e15}{\per\centi\meter\cubed}. 
Annealing in an NO environment therefore has two important implications. Firstly, defect passivation by N leads to a reduction of the interface trap concentration. The reduction of the anticipated Si vacancies near the interface region would result in an enhancement of the diamagnetic asymmetry signal according to the findings of Ref.~\cite{woerle2020muon}. 
Secondly, 
nitrogen introduced by NO annealing migrating through the oxide layer would create a region of high carrier concentration close to the interface if N occupies a substitutional lattice site in the SiC. The increased A$_\mathrm{D}$ signal for the NO annealed samples~(Fig.~\ref{fig:260K_5G}a) is therefore likely a result of both of these processes. 
The presence of residual V$_\mathrm{Si}$ after NO annealing is still possible but may be overshadowed in the muon data by the large increase in carrier concentration. 

\begin{figure}
	\centering
	\includegraphics{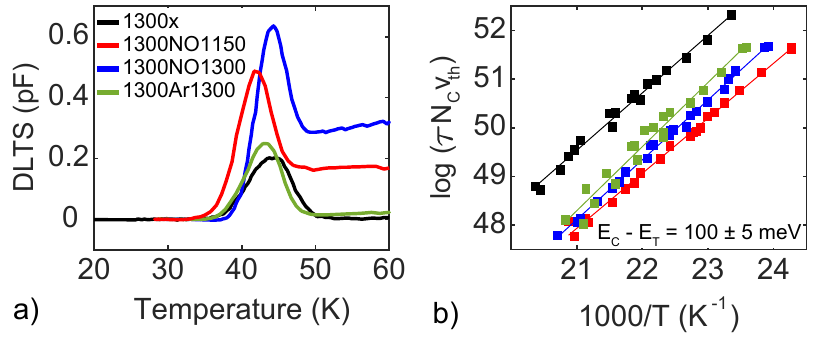}
	\caption{ DLTS spectra for a 0~V reverse bias and a pulse voltage of -5~V. a) Peaks between 40-45~K are recorded and correspond to an energy level of 100~$\pm$~5~meV below the conduction band edge as shown in b). This defect is associated with the N donor level at the cubic site. The peak amplitude is higher for the NO annealed samples indicating the increased N concentration due to the NO annealing.}
	\label{fig:DLTS}
\end{figure}

\begin{table*}[t]
\caption{\label{tab:tableDLTS} Electrical parameters of the N-related level measured in DLTS including the energy level below the conduction band edge, the apparent capture cross section and the ratio of the capacitance transient and the reverse bias capacitance at the peak in the DLTS signal. }
\begin{ruledtabular}
\begin{tabular}{cccc}
\textbf{Sample Name} & \textbf{E$_\mathrm{C}-$E$_\mathrm{T}$~(eV)} & \textbf{Capture cross section~(cm$^\mathrm{2}$)}& \textbf{$\Delta$C/C$_\text{R}$} \\
\hline
1300x & 0.098 & $6.1 \times 10^{-14}$ & $6.7 \times 10^{-2}$ \\
1300NO1150 & 0.098 & $4.1 \times 10^{-13}$ & $9.6 \times 10^{-2}$\\
1300NO1300  & 0.104& $1.3 \times 10^{-12}$ & $1.4 \times 10^{-1}$\\
1300Ar1300  & 0.105& $7.2 \times 10^{-12}$ & $7.3 \times 10^{-2}$\\
\end{tabular}
\end{ruledtabular}
\end{table*}

\noindent
The hypothesis that the muon signal in Fig.~\ref{fig:260K_5G} is responding to an enhanced carrier concentration caused by the NO anneal depends on the nitrogen atoms being able to traverse the SiO$_2$ layer and migrate up to hundred nanometers into the SiC crystal. 
Fig.~\ref{fig:DLTS}~a) shows the recorded DLTS signal for the four samples 1300x, 1300Ar1300, 1300NO1150 and 1300NO1300. 
The donor level assigned to substitutional N in the pseudo-cubic configuration, N$_\mathrm{C}$(\textit{k}) \cite{gotzNitrogenDonorsSilicon1993,kimotoNitrogenDonorsDeep1995, pernotElectricalTransportType2001,kagamiharaParametersRequiredSimulate2004,gelczukOriginAnomalousBehavior2020,assmannFineStructureElectronic2021}, is observed for all the samples at around 100~$\pm$~5~meV below the conduction band edge as shown in Fig.~\ref{fig:DLTS}~b). 
Electrical parameters of this defect level for the four samples are shown in Table~\ref{tab:tableDLTS}. 
The DLTS peak height is directly related to the defect concentration and it can be seen that for the NO annealed samples the peak height is substantially higher than for the non-annealed and Ar-annealed samples, evidencing the incorporation of N into the SiC crystal during NO annealing. Furthermore, the fact that the N-peak intensity is higher after annealing in NO at \SI{1300}{\celsius} as compared to the \SI{1150}{\celsius} case 
is an indication that the diffusion of N into the SiC increases with temperature. 
Hence, the DLTS data demonstrates that nitrogen atoms penetrate the oxide layer for the post oxidation annealing temperatures employed herein. 

\begin{figure}
	\centering
	\includegraphics{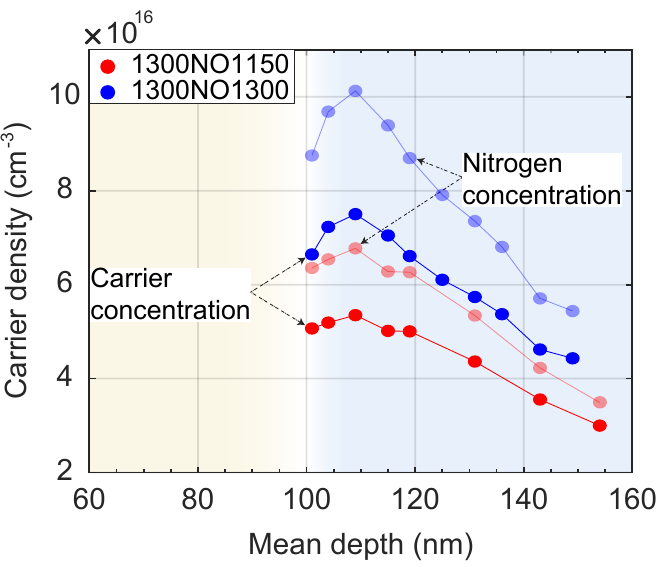}
	\caption{Extraction of N concentration and the effective carrier concentration near the oxide-semiconductor interface after the POA in NO using the LE-$\mu$SR data. The dark~(red and blue) curves show the concentration of carriers at 260~K and the semi-transparent curves show the concentration of N that must be present to have the calculated free carrier density. The doping concentration is increased by more than an order of magnitude for both the NO annealed samples compared to the background doping concentration of $\sim$\SI{2.6e15}{\per\centi\meter\cubed}. }
	\label{fig:CarrierConc}
\end{figure}

\subsection{Estimation of carrier concentration using LE-$\mu$SR}

\noindent The transition of Mu$^0$ to Mu$^-$ is dependent on the electron capture rate by the Mu$^0$ and is therefore proportional to the electron concentration which is equivalent to the concentration of ionized donors~($\text{N}_\text{D}^+$). Additionally, the process is also dependent on the electron capture cross section of the Mu$^0$~($\sigma_c^e$) and temperature~(T). With increasing temperature, the donor atoms are more easily ionized leading to an increase in 
electron concentration which supports the formation of Mu$^-$, however, the mobility of the electrons reduces leading to a reduction in the electron capture rate.
Using these parameters the diamagnetic asymmetry is modelled according to

\begin{eqnarray}
\text{A}_\text{D}\text{(T)} =  \text{c}_\text{1}\text{N}_\text{D}^\text{+}\text{$\sigma$}_\text{c}\left(\text{$\frac{\mathrm{T}}{300}$}\right)^{\text{c}_\text{2}} + \text{c}_\text{3}, 
\label{eq:AD_N+}
\end{eqnarray}

\noindent 
where c$_\mathrm{1}$, c$_\mathrm{2}$ and c$_\mathrm{3}$ are  fitting parameters.
A calibration was performed using
LE-$\mu$SR measurements on N-implanted 4H-SiC samples with a doping density of $1\times10^\mathrm{17}$~cm$^\mathrm{-3}$. The dopant densities were verified experimentally using secondary ion mass spectrometry (SIMS). More details are shown in the Appendix. \\
Having established the relation between the A$_\mathrm{D}$ and $\text{N}_\text{D}^+$ using the reference sample and identified the fitting parameters, Eq.~\ref{eq:AD_N+} can be used to estimate the $\text{N}_\text{D}^+$~(and N$_\mathrm{D}$) in narrow regions of a SiC sample by performing a LE-$\mu$SR measurement and extracting the A$_\mathrm{D}$ in a depth resolved manner. 
This method was employed for the two NO annealed samples 
and the evaluated N$_\mathrm{D}$ and N$_\mathrm{D}^+$ are shown in Fig.~\ref{fig:CarrierConc}. The dark red and blue curves represent the carrier concentration at 260~K, while the semi-transparent curves show the dopant concentration~(N$_\mathrm{D}$) near the interface. In both the NO annealed samples, a significant increase in the carrier concentration is observed compared to the background doping concentration of $\sim$2.6~$\times$~$\mathrm{10}^{15}~\mathrm{cm^{-3}}$. The high carrier concentration shows that N is able to diffuse into the SiC to concentrations as high as $1\times10^\mathrm{17}$~cm$^{-3}$ in the topmost tens of nanometer close to the SiO$_2$-SiC interface. The change in carrier concentration after NO annealing compared to the background doping persists up to at least \SI{50}{\nano\meter} in the SiC. \\
Due to the diffusion of N, the carrier concentration is not uniform and gradually decreases towards the SiC bulk. The impact of this gradual reduction can also be observed in the detector phase recorded for the 1300NO1300 sample in Fig.~\ref{fig:260K_5G}~d). As explained in the previous section, with reducing carrier concentration, the muon stays longer in the Mu$^0$ precursor state before transforming to Mu$^-$, leading to a more negative phase. \\
Diffusion of N to the SiO$_\mathrm{2}$-SiC interface also seems to have an impact on the oxide. Close to the interface, in the SiO$_\mathrm{2}$, the A$_\mathrm{D}$ signal~(see Fig.~\ref{fig:260K_5G}~a)) for the NO annealed samples appear to be slightly larger than for the Ar and the non-annealed samples. Presence of N in high concentrations at the interface could be responsible for this increment in the A$_\mathrm{D}$ signal.

\begin{figure}
	\centering
	\includegraphics[width=\columnwidth]{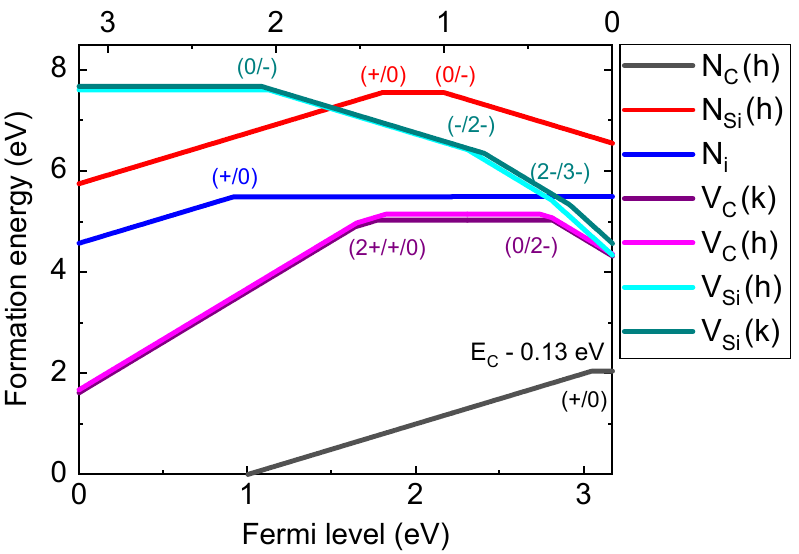}
	\caption{Formation energy diagrams for nitrogen and Si and C vacancies in 4H-SiC calculated using the HSE06 hybrid functional and 576-atom supercells.}
	\label{fig:ef-diag}
\end{figure}

\subsection{Theoretical defect kinetics}
Having established that nitrogen atoms from the NO anneal likely permeate the SiO$_2$-SiC interface and enter the silicon carbide, two questions remain: (i) where does the nitrogen sit in the 4H-SiC lattice, and (ii) is nitrogen migration in 4H-SiC feasible under the relevant conditions. Formation energy diagrams for various defect species were formed using density functional theory (DFT) calculations to address the first question. Fig.~\ref{fig:ef-diag} shows formation energy diagrams for the C and Si vacancies, N on C and Si lattice sites, and the nitrogen split-interstitial on a C lattice site. 
Note that only one lattice site configuration (hexagonal, \textit{h}, as opposed to pseudo-cubic, \textit{k}), was considered for the nitrogen related point defects herein.
The nitrogen split-interstitial was found to inhabit a neutral charge state in n-type and intrinsic 4H-SiC material. 

As also found in previous works \cite{deak1998,Rurali2003}, N on the C lattice site (N$_\mathrm{C}$) is more energetically favorable than N on the Si lattice site (N$_\mathrm{Si}$) by over \SI{5}{\electronvolt} in n-type 4H-SiC. Moreover, formation of N$_\mathrm{C}$ is energetically favorable as compared to V$_\mathrm{C}$, while this is only true for the Si lattice site in p-type material --- not in n-type. This can be understood in relation to the impact of the nitrogen on the surrounding material. N$_\mathrm{C}$ incurs little to no relaxation in the surrounding lattice.  
Indeed, nitrogen is similar to carbon in the fact that they have similar atomic radii being nearest neighbours on the periodic table and may exhibit $sp^\text{3}$-hybrid orbitals. The similarity between the electronic structures of N and C causes little additional energy cost upon replacing one with the other, while nitrogen cannot replace silicon in the Si-C crystal bonding structure.
Hence, N$_\mathrm{Si}$ causes a displacement of neighboring atoms that comes with an energy cost. 
In the negative charge state, N$_\mathrm{Si}^-$, the nearest neighbor C atoms to N$_\mathrm{Si}$ are distorted inwards in the basal plane while the axial C-atom experiences an outward relaxation along the $c$-axis. 
The dopant site competition for nitrogen in 4H-SiC is therefore expected to result in the vast majority of N atoms residing on the C lattice site~\cite{larkin_sitecompetition_1994}.

\begin{figure}
	\centering
	\includegraphics{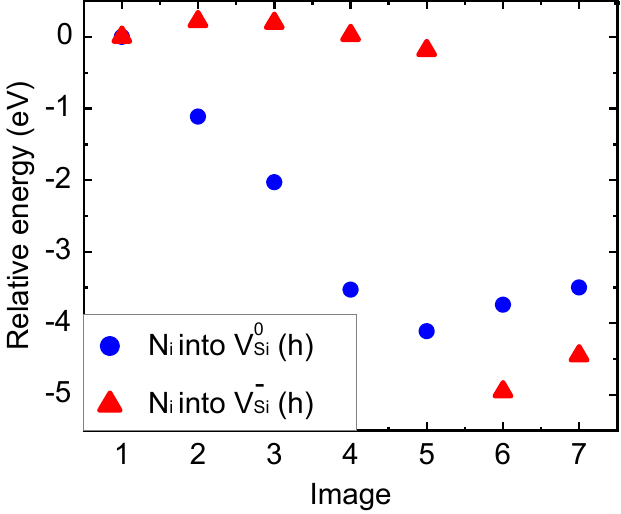}
	\caption{Minimum energy path calculated using the NEB method and PBE functional for the N$_\mathrm{i}$ entering the V$_\mathrm{Si}$ in the neutral (576-atom supercell, blue circles) and negative (400-atom supercell, red triangles) charge states. }
	\label{fig:mep-intoVsi}
\end{figure}

The LE-$\mu$SR data clearly demonstrates an increase in carrier concentration near the interface after NO annealing that was attributed to nitrogen donating electrons to the lattice. Fig.~\ref{fig:ef-diag} confirms that N$_\mathrm{C}$ acts as a shallow dopant and contributes electrons to the lattice. N$_\mathrm{Si}$, on the other hand, is found to act as a deep-level defect and charge carrier trap in both p- and n-type material. However, if we assume the presence of Si vacancies after thermal oxidation (as shown in, e.g., Refs.~\cite{cochraneIdentificationSiliconVacancy2012, cochrane2013effect}), the formation of N$_\mathrm{Si}$ can, in fact, free electrons to the lattice. The V$_\mathrm{Si}$ occupies the triply negative charge state in n-type material while N$_\mathrm{Si}$ is only single negatively charged. Hence, the transformation of V$_\mathrm{Si}$ into N$_\mathrm{Si}$ would release charge carriers to the lattice and contribute to an increase in the effective carrier density. 

The hypothesis presented in Fig.~\ref{fig:CarrierConc}, that nitrogen enters the 4H-SiC lattice to a range of some hundred nanometer, relies on nitrogen being able to migrate in 4H-SiC at \SI{1300}{\celsius} during the NO anneal.  
Previous theoretical studies find activation energy barriers for migration of interstitial nitrogen in the neutral charge state in the \SIrange{2.6}{3.0}{\electronvolt} range \cite{Gerstmann2003, rauls_different_2003}. Assuming that the diffusivity of N$_\mathrm{i}$ is given by $D=D_0 \exp (-E_\mathrm{A}/k_\mathrm{B}T)$, the diffusion length is given by $l=\sqrt{D\tau}$, and the pre-factor for diffusion is $D_0 \sim 0.001$~cm$^2$s$^{-1}$ \cite{philibert1991}, the diffusion length for N$_\mathrm{i}$ ranges from around \SI{300}{\nano\meter} to \SI{1.4}{\micro\meter} for activation energies of 3.0~eV and 2.6~eV, respectively, for an NO anneal at \SI{1300}{\celsius} for \SI{70}{\min} (see Table~\ref{tab:table1}). Hence, the migration of N within the 4H-SiC is deemed feasible herein during the POA in NO --- provided that the N atoms are able to traverse the barrier between the SiO$_2$ and 4H-SiC, the strength of which is unknown. 

The higher energy cost for N to occupy a Si site as compared to a carbon site leaves the question of whether N$_\mathrm{Si}$ defect formation is at all feasible. Assuming that a N atom sits in an interstitial site close to the V$_\mathrm{Si}$, we find that the defect complex N$_\mathrm{Si}$ is lower in energy by as much as \SIrange{3.7}{4.2}{\electronvolt} depending on the charge state. In this case, where the V$_\mathrm{Si}$ and the N$_\mathrm{i}$ are both already present and in close proximity, the formation of N$_\mathrm{Si}$ is clearly energetically favorable. 

Fig.~\ref{fig:mep-intoVsi} shows the migration energy paths of N$_\mathrm{i}$ entering the neutral V$_\mathrm{Si}$(\textit{h}) within the basal plane (blue circles) and the negative V$_\mathrm{Si}$(\textit{h}) along the axial direction (red triangles). For both cases there is a substantial energy gain for the transformation to occur, and there is no additional energy barrier for the transformation N$_\mathrm{i}+$V$_\mathrm{Si}\rightarrow$ N$_\mathrm{Si}$ for the negative nor the neutral charge state. Note that the defect geometries along the pathways with energies lower than the final structure are an artificial result of the use of the $\Gamma$-point only for \textbf{k}-point sampling for the NEB calculations, while a $2\times2\times2$ \textbf{k}-mesh was employed for the original structural relaxation. 

The results discussed so far demonstrate that (i) N on both the C and Si lattice sites can contribute to the increase in carrier density observed after NO annealing, (ii) N migration in 4H-SiC is feasible under the studied POA conditions, and (iii) the formation of N on Si site is possible provided that both N$_\mathrm{i}$ and V$_\mathrm{Si}$ are present and in close proximity. 
The other possible mechanism for nitrogen to provide electrons to the lattice is that the migrating N$_\mathrm{i}$ spontaneously kicks out a C atom from a lattice site according to the reaction pathway N$_\mathrm{i}+\mathrm{C}_\mathrm{C}\rightarrow \mathrm{N}_\mathrm{C}+\mathrm{C}_\mathrm{i}$. This mechanism was not studied herein but presents an  alternative pathway to the formation of N$_\mathrm{Si}$ considering the much larger energy gain for N to sit on a C compared to a Si lattice site (see Fig.~\ref{fig:ef-diag}).  Ref.~\cite{Gerstmann2003} considered a related but slightly different reaction, where instead of occupying the V$_\mathrm{Si}$, the interstitial N on a split-interstitial C site kicks out a C atom to form the nitrogen-vacancy center and a C split-interstitial: 
N$_\mathrm{i}+\mathrm{V}_\mathrm{Si}\rightarrow \mathrm{N}_\mathrm{C}\mathrm{V}_\mathrm{Si}+\mathrm{C}_\mathrm{i}$. 
The corresponding relatively low activation energy of \SI{2.9}{\electronvolt} is comparable to that expected for N$_\mathrm{i}$ migration and suggests that the spontaneous formation of N on the C site is a potential candidate for the observed high carrier densities near the SiC-SiO$_2$ interface (see Fig.~\ref{fig:CarrierConc}).

\begin{figure}
	\centering
	\includegraphics{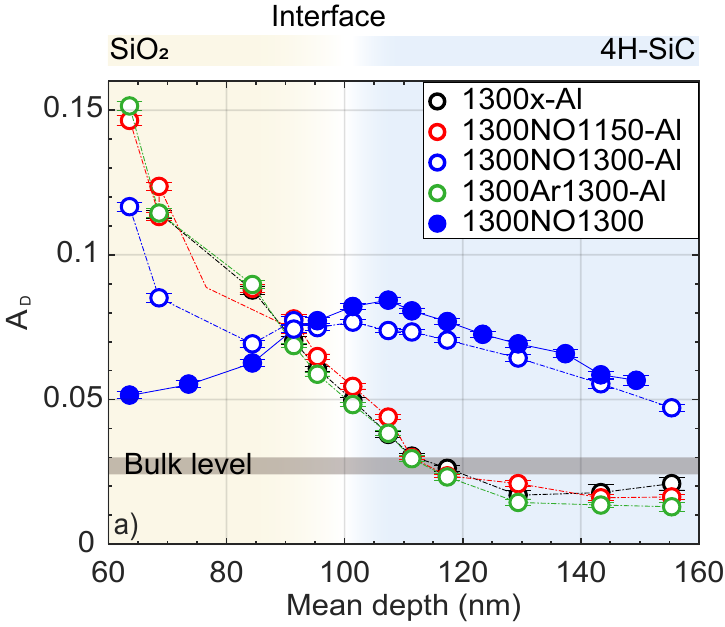}
	\caption{Diamagnetic asymmetry A$_\mathrm{D}$ of the MOS samples at 260~K and 0.5~mT as a function of mean depth.
	A depletion region is created for the non-annealed, Ar annealed and the NO annealed sample at 1150~$^\circ$C resulting in a reduced A$_\mathrm{D}$ signal in the first 60~nm of SiC. 1300NO1300-Al has a near flat-band condition and shows only a slight reduction in the A$_\mathrm{D}$ signal.}
	\label{fig:Ad_withAl}
\end{figure}

\subsection{LE-$\mu$SR study of metal-oxide-semiconductor~(MOS) samples}

\begin{figure}[t]
	\centering
	\includegraphics{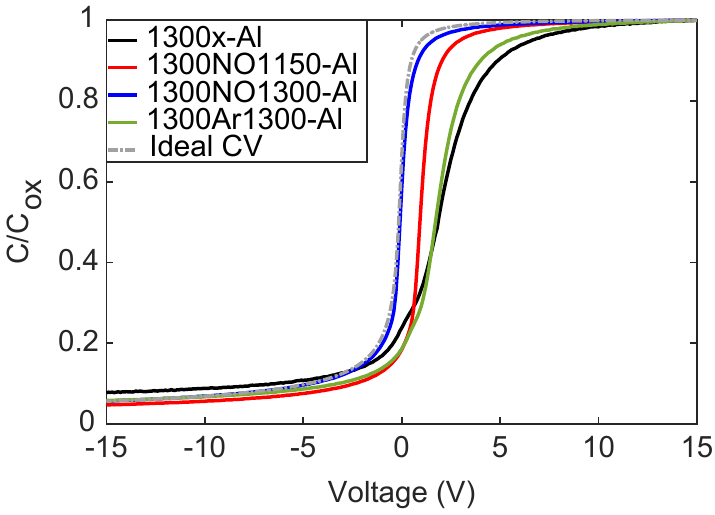}
	\caption{Normalised high frequency CV curves for the MOS samples in comparison to the ideal C-V curve. The normalization is performed by using the oxide capacitance~(C$_\text{OX}$) of the MOS capacitor. 
	}
	\label{fig:CV}
\end{figure}

\begin{table*}[t]
\caption{\label{tab:tableMOS} Electrical parameters of the MOS capacitors.}
\begin{ruledtabular}
\begin{tabular}{cccc}
Sample Name & Flat-band voltage~(V$_\mathrm{FB}$)&Surface potential at zero bias~($\Psi_\mathrm{s0}$)& Depletion width~(W$_\mathrm{D}$) \\
\hline
1300x-Al &2.3~V& -0.896~V & 430~nm\\
1300NO1150-Al &1.1~V & -0.420~V & 294~nm\\
1300NO1300-Al &0.1~V& -0.034~V & 84~nm \\
1300Ar1300-Al &2.0~V& -0.781~V & 401~nm 
\end{tabular}
\end{ruledtabular}
\end{table*}

\noindent In a next step, we investigated the impact of adding a thin layer of metal~(Al) on top of the thermal oxide in our samples. 
The LE-$\mu$SR data for the MOS samples at 260~K and 0.5~mT are shown in Fig.~\ref{fig:Ad_withAl}.
The A$_\mathrm{D}$ signal for all samples except for the sample annealed in NO at 1300~$^\circ$C falls below the bulk level for the MOS capacitor case at mean depths $> 120$~nm, i.e. about $\sim 20$~nm away from the interface.
Interestingly, the A$_\mathrm{D}$ values for the 1300x-Al, 1300NO1150-Al and 1300Ar1300-Al MOS samples at 260~K are very similar to those recorded for the SiO$_2$-SiC samples at 10~K~(see Fig~\ref{fig:10K_5G}). While for the oxide-semiconductor samples, a low A$_\mathrm{D}$ at 10~K was attributed to the lack of charge carriers due to low temperature, a drop in A$_\mathrm{D}$ for the MOS samples at 260~K could be due to the creation of a depletion region near the interface. 
\\
C-V measurements were performed in quasi-static mode and at high frequency to understand the origin of this alleged depletion region and to extract the density of interface traps. Fig.~\ref{fig:CV} shows the normalized high frequency capacitance data for the  samples. The C-V data for 1300NO1150 is laterally shifted when compared to the 1300NO1300 data but not stretched, indicating the presence of a higher density of fixed oxide charges but a  similar level of D$_\mathrm{it}$. This is also seen in the extracted D$_\mathrm{it}$ shown in Fig.~\ref{fig:Dit}, where the defect densities are very similar for the two NO annealed samples near the conduction band edge, and differ only beyond E$_\mathrm{C}-$0.3~eV. The absence of a lateral shift for the 1300NO1300 sample in comparison to the ideal CV curve also indicates that the NO annealing at 1300~$^\circ$C passivates the fixed oxide charges. The C-V data for the 1300x sample is further shifted to the right in comparison to the 1300NO1150 sample and is also stretched out indicating a higher density of fixed oxide charges and interface traps when compared to the NO annealed samples. In comparison, the Ar annealed sample has a steeper CV curve but a similar lateral shift. This indicates that annealing in an Ar environment helps reduce the D$_\mathrm{it}$~(see Fig.~\ref{fig:Dit}) but the non-annealed and the Ar annealed samples have similar concentration of oxide charges. 
\\
The fixed oxide charges have an impact on the flat-band voltage and consequently on the surface potential at zero bias for the MOS capacitors~\cite{schroderOxideInterfaceTrapped2005}. The flat-band capacitance of the semiconductor is calculated using the Debye length and is then used to extract the flat-band voltage~(V$_\mathrm{FB}$) of the MOS structures. The surface potential and the depletion region width at zero bias~($\Psi_\mathrm{s0}$, W$_\mathrm{D}$) are extracted using the V$_\mathrm{FB}$ and are shown in Table~\ref{tab:tableMOS}. 
The $\Psi_\mathrm{s0}$ values for the 1300x-Al, 1300NO1150-Al and 1300Ar1300-Al samples are significantly higher than for the 1300NO1300-Al sample. This translates directly to a large depletion region width near the interface which would explain the drop in the A$_\mathrm{D}$ signal for these samples. For the 1300NO1300-Al sample, a very small $\Psi_\mathrm{s0}$ implies a near flat-band condition. A depletion region width of 84~nm is calculated for the 1300NO1300-Al sample considering a background doping concentration of \SI{2.6e15}{\per\centi\meter\cubed}. Accounting for an increase in the carrier concentration by more than one order of magnitude to \SI{7e16}{\per\centi\meter\cubed} in the topmost 60~nm beneath the SiO$_2$-SiC interface, the depletion region width is expected to drop to $\sim$20~nm. A small drop in A$_\mathrm{D}$ in the first 20~nm recorded for the 1300NO1300-Al sample (see Fig.~\ref{fig:Ad_withAl}) aligns very well with the C-V measurement and the predicted depletion region width. \\
The measurements on the MOS capacitors demonstrate that the variation in charge carrier concentration due to an external electric field can be studied with LE-$\mu$SR, opening doors for a detailed and unprecedented study of accumulation, depletion and inversion in the near interface region of a metal oxide semiconductor field effect transistor~(MOSFET) with nano-meter depth resolution.

\section{Concluding remarks}
In this work, we have explored the use of LE-$\mu$SR to understand the impact of thermal oxidation of SiC and post oxidation annealing in an Ar and NO environment in a depth-resolved manner. With excellent control over the implantation depth of muons, a study of the SiO$_2$-SiC interface and near interface region has been performed with nanometer-depth resolution. A comparison of low and high temperature measurements reveal that a carrier rich region is formed near the interface after NO annealing extending up to at least 50~nm into the SiC. Building upon our previous experiments with $\mu$SR and the known affinity of $\mu^+$ to form the V$_\mathrm{Si}$-Mu$^0$ complex, the increase in A$_\mathrm{D}$ for the NO annealed samples points toward a reduction in the concentration of V$_\mathrm{Si}$ due to passivation by N and due to an increased electron concentration. Since a significant drop in the D$_\mathrm{it}$ is also observed for the NO annealed samples, we speculate that V$_\mathrm{Si}$ is a major contributor to the defects at the SiO$_\mathrm2$-SiC interface. \\
A quantitative analysis of the carrier concentration near the interface was demonstrated using the diamagnetic signal extracted from the LE-$\mu$SR measurements. By controlling the probing depth of the muons, small variations in the carrier concentration due to accumulation or depletion in the material can be studied. A doping concentration of 1~$\times$~$\mathrm{10}^{17}~\mathrm{cm}^{-3}$ is estimated close to the interface for the sample annealed at 1300~$^\circ$C in NO environment which drops quickly upon moving into the bulk of SiC. An increase in the N concentration is also observed in the DLTS measurements. Variations in the shallow N level across the sample set indicate that the N concentration is substantially higher (by a factor of 2-3) for the NO annealed samples when compared to the non-annealed sample. \\
Density functional theory calculations performed in this work and the literature corroborate the possibility for nitrogen migration some 100~nm in 4H-SiC during the POA in NO. Both N on C and Si sites are found to contribute electrons to the lattice upon defect formation. Hence, two possible pathways are found: interstitial N encounters a Si vacancy and forms the N$_\mathrm{Si}$ deep level defect, or migrating N$_\mathrm{i}$ spontaneously kicks out a C atom from the lattice to form the N$_\mathrm{C}$ shallow dopant and a C split-interstitial. 
\\
Furthermore, the impact of depositing a metal on the oxide-semiconductor structure to form a MOS capacitor is investigated using LE-$\mu$SR and C-V measurements. We observe that a depletion region is created near the interface which results in a drop of A$_\mathrm{D}$ for all the samples. While only a slight reduction is observed for the 1300NO1300 sample due to a narrow depletion region, the A$_\mathrm{D}$ for the other samples is reduced below the bulk value indicating a much broader space charge  region. \\
Importantly, with the proper calibration available, LE-$\mu$SR proves to be a powerful tool for understanding device interfaces which are known to impact device operation. Despite the limitations and time constraints for employing the technique, we expect muon spin rotation to provide important insights into the carrier, dopant and defect concentrations and distributions in the near interface region and hence to provide valuable information for the further development of SiC power devices.
 
\section*{Acknowledgements} 
The muon experiments were performed at the $\mu$E4/LEM beamline \cite{Prokscha_2008} of the Swiss Muon Source S$\mu$S, Paul Scherrer Institute, Villigen, Switzerland. This work is supported by the Swiss National Science Foundation under the Grant No. 192218. The work of MEB was supported by an ETH Zurich Postdoctoral Fellowship. The computations were performed on resources provided by UNINETT Sigma2 -- the National Infrastructure for High Performance Computing and Data Storage in Norway.  

\section{Appendix: Electron capture process for the formation of the diamagnetic signal in a N-implanted sample. } 

\begin{figure}[h]
	\includegraphics{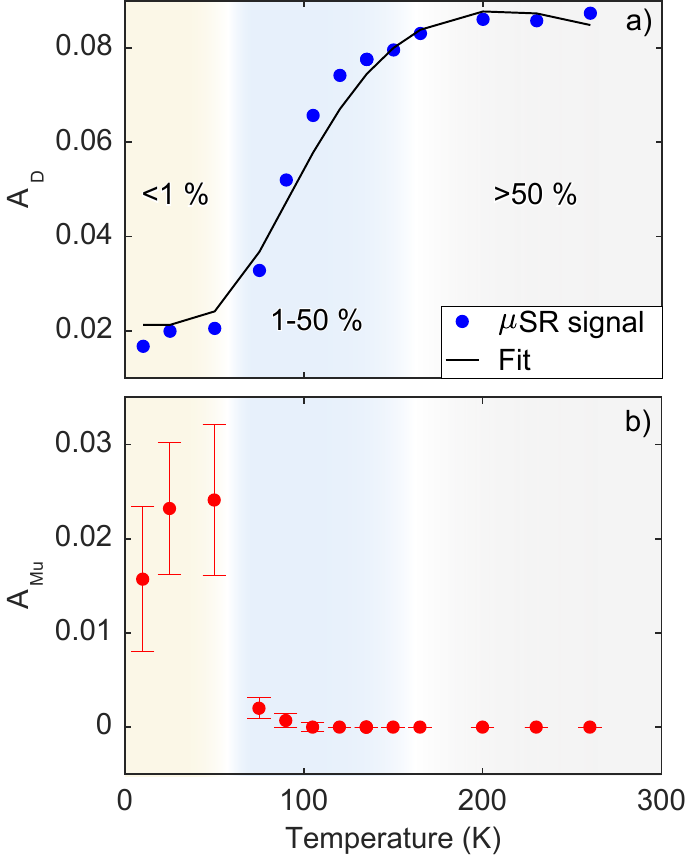}
	\caption{a) Diamagnetic asymmetry A$_\text{D}$ as a function of temperature T of a 10$^{17}$~cm$^{-3}$ N-implanted sample. The solid line is a fit of Eq.~3 to the data.
	Three zones of temperature are shown in the figure with low~($< 1\%$), medium~(1-50~$\%$) and high($> 50\%$) degree of donor ionization b) Paramagnetic asymmetry A$_\text{Mu}$ is high in the low temperature range but drops quickly to almost zero as the temperature rises beyond 100~K. }
	\label{fig:AlphaCalibration}
\end{figure}
In a moderately to highly doped n-type 4H-SiC sample, the Mu$^0$ formed at the end of the ionization track can capture an electron from the semiconductor to form the diamagnetic state Mu$^-$. Prokscha~${et~al.}$~\cite{prokschaDirectObservationHole2020a,prokschaSimulationTFmSRHistograms2014a} showed that for Ge, the capture of holes by the muonium depends on the concentration of holes, the capture cross section~($\sigma_c^h$) and the thermal velocity~(v$_h$) of the carriers. It was further found that the product of v$_h$ and $\sigma_c$ was proportional to T$^{-2.4}$, indicating that the hole capture rate by muonium is governed by the mobility of the holes. 
Using a similar approach for n-type samples, the transition of Mu$^\mathrm{0}$ to Mu$^\mathrm{-}$ would also be directly proportional to the density of free electrons~($\sim \text{N}_\text{D}^+$), the electron capture cross section of the Mu$^\mathrm{0}$~($\sigma_c^e$) and the temperature. With increasing temperature, more donor atoms are ionized supporting the transition to the diamagnetic state, but at the same time, the mobility of carriers is reduced, 
counter acting the electron capture rate increase due to the increasing electron concentration.
In the low temperature range~($<$60~K), the electron density is very low and the ionization is smaller than 1~\% of the doping density. Consequently, A$_\mathrm{D}$ is very small and a large fraction of the implanted muons exist in the paramagnetic state as shown in Fig.~\ref{fig:AlphaCalibration}~a, b). As the temperature is increased, the electron density increases and the transition of Mu$^\mathrm{0}$ to Mu$^\mathrm{-}$ is supported. This is evident from the blue region in the Fig.~\ref{fig:AlphaCalibration} where a rapid rise in A$_\mathrm{D}$ at the cost of A$_\mathrm{Mu}$ is observed. Beyond 180~K, more than 50~\% of the donors are ionized and the sample has a high electron density. However, the increasing temperature also adversely affects the electron capture rate of Mu$^\mathrm{0}$, reducing the rate of increase of diamagnetic asymmetry with temperature.\\ 
Further, to calculate the N$_\mathrm{D}^+$ in our reference sample, the information of the donor levels is crucial. N-related donor levels in 4H-SiC have been reported in the literature and range from 40-65~meV for the hexagonal site and 85-125~meV for the cubic site  \cite{gotzNitrogenDonorsSilicon1993,kimotoNitrogenDonorsDeep1995, pernotElectricalTransportType2001,kagamiharaParametersRequiredSimulate2004}. In our analysis we have chosen two donor levels within this range: 50~meV~(hexagonal site) and 100~meV~(cubic site), with each level occupying half of the dopants. This assumption has a certain error margin due to the spread in the reported values, and a variation in the ratio of N$_\mathrm{C}$(\textit{h}) to N$_\mathrm{C}$(\textit{k}) may lead to discrepancies in the calculated carrier concentration. Accounting for the spread of the energy levels from various sources in literature, a variation of up to 25~\% can occur in the calculated carrier concentration when comparing the lower and upper bounds for the N-related donor levels: N$_\mathrm{lower}$~(40~meV, 85~meV) and N$_\mathrm{upper}$~(65~meV, 125~meV).\\


\noindent
Taking into account the impact of the ionization of dopants~($\text{N}_\text{D}^+$) and the temperature, A$_\mathrm{D}$ can be modelled for the entire temperature range according to the  Eq.~\ref{eq:AD_N+} with calibration coefficients c$_\mathrm{1}$, c$_\mathrm{2}$ and c$_\mathrm{3}$. The $\text{A}_\text{D}$ signal for the 10$^{17}$~cm$^\mathrm{-3}$ N-implanted sample is used for calibration. The muon implantation energy was kept at 18~keV which would correspond to a mean depth of $\sim$100~nm in the SiC. The sample underwent a post implantation annealing at 1700~$^\circ$C for thirty minutes and was then thoroughly cleaned in 2\% HF solution to remove any native oxide before the measurement.\\
Assuming a constant capture cross section~($\sigma_c^e$) of 1$\times$10$^{-15}$~cm$^{2}$, and calculating the N$_\mathrm{D}^+$ at each temperature~(100\% donor activation is assumed), Eq~\ref{eq:AD_N+} is solved to get the best fit of the acquired data. The extracted calibration coefficients are c$_\mathrm{1}$ = 7.08~$\times$10$^\mathrm{-4}$, c$_\mathrm{2}$~=~-1.33 , c$_\mathrm{3}$~=~0.021. \\
The recorded A$_\mathrm{D}$ and the fit is shown in the Fig.~\ref{fig:AlphaCalibration}~a). 
Using these coefficients a very good agreement between the experimental data and the model (Eq. \ref{eq:AD_N+}) is observed. \\

\normalem
\bibliography{apssamp}

\end{document}